*Article*

# A Comparative Analysis of Bias Amplification in Graph Neural Network Approaches for Recommender Systems

Nikzad Chizari , Niloufar Shoeibi  and María N. Moreno-García * 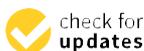

Department of Computer Science and Automation, Faculty of Sciences, University of Salamanca, Plaza de los Caídos sn, 37008 Salamanca, Spain
* Correspondence: mmg@usal.es

**Abstract:** Recommender Systems (RSs) are used to provide users with personalized item recommendations and help them overcome the problem of information overload. Currently, recommendation methods based on deep learning are gaining ground over traditional methods such as matrix factorization due to their ability to represent the complex relationships between users and items and to incorporate additional information. The fact that these data have a graph structure and the greater capability of Graph Neural Networks (GNNs) to learn from these structures has led to their successful incorporation into recommender systems. However, the bias amplification issue needs to be investigated while using these algorithms. Bias results in unfair decisions, which can negatively affect the company's reputation and financial status due to societal disappointment and environmental harm. In this paper, we aim to comprehensively study this problem through a literature review and an analysis of the behavior against biases of different GNN-based algorithms compared to state-of-the-art methods. We also intend to explore appropriate solutions to tackle this issue with the least possible impact on the model's performance.

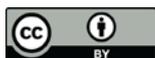

## 1. Introduction

Currently, numerous users benefit from the advantage of purchasing different products and services online. The problem arises when people face too many options, which can cause an overload of information, leading to a difficult decision-making process. To overcome this problem, Recommender Systems (RSs) are used to provide users with personalized item recommendations [1–5].

Considering the vast usage of RSs in a wide range of application domains, it is important to make sure the recommendation results are not unfair. Bias is one of the most important concerns in RSs, which affects RSs' effectiveness [1,2]. Biased recommendation results can cause an unbalanced and unfair allocation of resources and opportunities [6]. These biased decisions can lead to severe financial, societal, and reputational harm to individuals and companies. Furthermore, bias can result in discrimination, which is against regulations [7–9]. Bias also can cause severe legal, technological, and security harms [10]. The mentioned problems have opened the door to the investigation of the bias issue in RSs in recent years, and the number of articles addressing this topic has increased [2]. These problems derived from bias in RSs are the main motivation of this article.

Lately, deep learning methods also have been improved significantly and found a path to RSs. Deep learning methods are capable of establishing a multi-layer, nonlinear, layer-to-layer interconnection network structure that aims to automatically extract representations of multi-level features from data [11]. These methods can significantly enhance the RSs' performance and address their problems. Among these methods, Graph Neural Network (GNN) algorithms have proven their usefulness in many learning tasks that require handling graph data, which contain rich information about relations between elements. GNNs can seize the dependence of graphs by propagating the information through the graph edges via message passing [12].

In order to address the limitations of traditional RSs, GNN algorithms have been successfully applied to RSs. The use of GNNs in multiple application domains has spread rapidly in recent years. This is mainly due to their power to learn from graphical information representations and the importance of the advantages of deep learning methods.

In recommender systems, user–item interactions can be modeled as graphs. Besides, additional data can be used to improve recommendations, including social or contextual information. In the RS field, neural-network-based methods, especially those using deep learning algorithms, have been proposed as an alternative to Collaborative Filtering (CF) approaches, thanks to their power to learn complex relations between users and items. However, these methods can only operate on Euclidean space data since they are not designed to deal with high-order information structures [6,12]. These drawbacks can be addressed by recent GNN techniques, which extend deep learning algorithms to the non-Euclidean space [13].

In GNNs, an iterative process is performed, where the information between nodes is propagated and aggregated from neighbors. So far, there are several proposals for GNN-based recommender models, both general and focused on sequential or session-based recommendations [14]. Different GNN variants can be used to accomplish these tasks. Among them, Graph Convolutional Networks (GCNs), Graph Attention Networks (GATs), and Graph Recurrent Networks (GRNs) have shown higher performance than other machine learning methods in a wide variety of areas [12], including recommender systems. Despite the proclaimed strengths of GNN-based models, there are still some challenges to be addressed for these approaches to prevail over the classical methods. One of them is dealing with multiple types of biases that affect recommender systems negatively. Some studies have evidenced that fairness and other social bias could be amplified by the use of graph structures [15].

Generally, bias in RSs can be categorized into four main classes according to Baeza-Yates (2016) and Chen et al. (2020) [2,16]: data bias, algorithmic bias (model bias), result and response bias, and feedback loop bias, which can magnify this issue. All of them can be divided into multiple categories. Bias also can cause serious problems in different aspects: economic, legal, social, security, and technological [17], as well as ethical aspects including content, privacy, autonomy, opacity, fairness, and social aspects [18,19]. In addition, objectives are influenced by bias regarding utility, coverage, diversity, novelty, serendipity, visibility, and exposure [20]. Bias also can be conveyed by user reviews such as sequential bias, opinion bias, and textual bias [21,22]. Bias can have different types o platforms, which can consist of functional biases, normative biases, external biases, and non-individual accounts [23]. This high impact of biases in multiple fields of study and application domains highlights the importance of analyzing them in-depth and proposing effective ways to address this problem. Biases in the RS area have been studied for several years. However, their study in the context of new algorithms such as GNNs is very limited.

In this paper, we aim to analyze the classification of the different types of biases troubling GNN-based RSs in general. We also analyze in depth the behavior of GNN algorithms against different types of biases (considering there is limited research analyzing the behavior of these algorithms against certain types of biases) and compare them with other widely used algorithms. We investigate if, in addition to achieving high accuracy in recommendation results, it is possible to trust that the algorithm is not biased and does not make unfair decisions. We also look into which biases are more accentuated with GNN algorithms and study ways of dealing with them.

In the following section, we present the state-of-the-art of the bias problem in Machine Learning (ML), GNN algorithms, and RSs, and finally, we focus on the problem of bias amplification in GNN-based RSs. In this section, we consider the most important related works in this specific area of research. Then, we explain the experimental study in detail. The results of the comparative study involving traditional and GNN-based recommenda-tion methods and two real datasets are presented in Section 4. In the end, we discuss the most important problems in this field and our future work.

## 2. State-of-the-Art

In this section, a review of the bias problem from different aspects including machine learning, GNN algorithms, RSs, and GNN-based RSs is presented. There is extensive work in the literature studying biases from different perspectives. However, in this section, we also analyze the most important works regarding the bias problem in general, focusing mainly on RSs and more specifically on GNN-based RSs.

### 2.1. Bias in Machine Learning

In order to have a better understanding of the bias problem regarding different aspects and how this problem can be transferred to the results of a recommender system, here, we provide different definitions of bias at different levels such as the statistical meaning of bias and bias in Machine Learning (ML).

According to [24], statistical bias relates to systematic errors produced by the measurement or sampling process. Within them, it is necessary to differentiate between errors caused by random chance and those caused by bias.

Data are the input in ML algorithms. When this input is biased, this bias can be transferred into the model generated by the algorithm, leading to unfair decisions and a reduction in quality [25,26]. These problems can have severe financial, social, and reputational effects on companies [8].

Algorithmic bias in ML-based models stems from abnormal datasets, weak models, poor algorithm designs, or historical human biases [27]. Algorithmic bias also can happen due to the problem of under-fitting in the training phase, which can be caused by a mixture of limitations in the training data and model capacity issues [28]. The factors affecting this mechanism are irreducible error (Bayes error), regularization mechanisms, class imbalance, and under-represented categories [28].

There are various types of general biases happening in different stages of CRISP-DM, a well-known standard process for data mining introduced by IBM in 2015, which breaks the data process into six different stages including business understanding, data understanding, data preparation, modeling, evaluation, and development. The types of described biases are social bias, measurement bias, representation bias, label bias, algorithmic bias, evaluation bias, deployment bias, and feedback bias [8]. Social bias happens when the data transfer biases in society to the model on a large scale. Measurement bias arises due to human error in the business understanding phase, especially in working with sensitive features such as age and gender. This kind of bias can also happen during the data preparation phase. Representation bias, moreover, occurs during data collection and sampling when the sample or data distribution does not represent the real underlying distribution. Label bias can be seen during the data preparation phase when labels are chosen for the prediction task. Choosing the best label for a dataset could be very difficult due to vagueness and cultural or individual variety. Algorithmic bias can also happen in the modeling phase due to model parameters or technical issues such as model misclassification [29].

It is also important to know, based on training data statistics, whether a model can amplify existing bias in data [29]. Evaluation bias, which happens in the evaluation phase, can happen because of the differences between the training and test data population. Finally, deployment bias can happen after model implementation in a complicated socio-technical environment.

Bias in ML can also lead to unfair results. Fairness in machine learning can be categorized into ten classes: statistical party, equalized odds, equal opportunity, disparate impact, disparate mistreatment, treatment equality, general entropy index, individual fairness (formerly, fairness through awareness), fairness through unawareness, and counterfactual fairness [30].

A systematic, controlled study on bias amplification is provided in [29]. To reach this, a heavily controlled simple image classification problem was taken into consideration. The results showed different factors including the accuracy of the model, model capacity, model overconfidence, and size of the training data are correlated with bias amplification. Furthermore, The results also illustrated that bias amplification can be different during training time, and also, the difficulty of classification tasks in recognizing group membership can influence bias amplification.

### 2.2. Bias in GNNs

GNNs and their variants have shown great performance for a wide range of graph learning tasks. However, they face remarkable computational challenges due to the increasing sizes of current datasets. Graph convolutions' multilayers mean recursively developing the neighbor aggregation in a top-down method, which can lead to a neighborhood whose size is growing based on the number of layers. If the graph is scale-free and condensed, a large part of the graph is required to compute the embeddings, also with a few layers, which is infeasible for large-scale graphs [31,32].

Other researches showed that GNNs perform better with homophilous nodes rather than heterophilous ones. A homophily ratio is defined in order to examine whether a graph is homophilous or heterophilous. Graphs with higher homophily ratios are considered homophilous, and graphs with lower ratios are non-homophilous [33].

Although GNNs usually provide better accuracy in results, most of the existing GNNs do not take the fairness issue into consideration, which can result in discrimination toward certain demographic subgroups with specific values of features that can be considered sensitive, such as age, gender, and race. The decision made by the implemented GNNs can be highly affected by these kinds of discrimination [32,34,35]. In addition, a wide range of ML systems are trained with human-generated data; hence, there is a clear need to comprehend and mitigate bias toward demographic groups in GNN approaches [36].

Biased results in GNN algorithms can stem from different reasons, the most important of which is a biased network structure. Although it is very important to detect which part of this network structure can lead to bias, it is believed this bias can be due to the message passing mechanism in the GNN's main operation. There are several challenges to understanding bias in the network structure, including the Fairness Notion Gap, Usability Gap, and Faithfulness Gap. The Fairness Notion Gap points to how to measure bias at the instance level. The Usability Gap points to the fact that it is also vital to find the edges in the computational graph most influential on the fairness degree of its prediction. The final edges cannot be considered the ones that contributed the most to this fairness. The Faithfulness Gap points to the need to ensure that gathered bias explanations indicate the true reasoning results based on the chosen model [34]. Furthermore, bias can also lead to a distribution shift between training and testing data, especially among labels [37].

### 2.3. Bias in RSs

The quality of recommendations provided by different RSs is various for different users based on their characteristics and sensitive information including age, gender, race, and personality. This behavior conflicts with European Commission (EC) regulations: "obligations for ex-ante testing, risk management and human oversight of AI systems to minimize the risk of erroneous or biased AI-assisted decisions in critical areas such as education and training, employment, important services, law enforcement, and the judiciary" [7]. According to this regulation, AI systems should follow EU fundamental rights such as the right not to be discriminated against, respecting individuals' private life, and personal data protection [7]. Moreover, biased results in RSs can cause user dissatisfaction [38].

Considering the bias issue in RSs is one of the most important factors leading to unfair decisions and discrimination, and this issue clearly disagrees with the mentioned regulations. The work presented in [2] indicates that bias can be divided into three potential categories, which can be the first considered for recognition: bias in input data, computational bias, which may stem from the algorithm and can be added to team decisions, and outcome bias. This is an expansion of the bias categorization previously introduced by Baeza-Yates (2016) [16], which helps break down the circular behavior into seven different types of biases in a circular format. Data bias, which is observational rather than experimental, happens when the distribution of the training data differs from the ideal test data distribution and consists of: selection bias, exposure bias, conformity bias, and position bias. Algorithmic bias can also happen during the different stages of the modeling process including training, evaluation, and feature engineering. Popularity bias, unfairness, and inductive bias can be the results of this particular type of bias. Popularity bias stems from the long-tail phenomenon in RSs. This common issue happens when a small number of very popular items have the most interaction in the system. This can lead to a neglection of the model toward unpopular items and give a higher score to the more popular ones [2]. Together, the previously mentioned biases can create a circle graph in which biased data move from one stage to the next, where additional and new biases are introduced [2,39]. This circular behavior of biases increases the complexity to recognize where actions are needed. Exposure bias happens due to the exposure of specific parts of items for users achieved from implicit feedback, and it can also be caused by popularity bias due to the recommendation of the most popular items [2,40]. In other words, bias can result in the limitation of the users' choices and contaminate users' feedback, which can amplify exposure bias [41]. Among the mentioned types of biases, popularity bias has been considered the most important type in the field of RSs [42].

Another proposed classification by Ashokan and Hass (2021) [30] for bias considers three main categories in more detailed sub-categories consisting of: data generation bias, which includes historical, representation, measurement, population, sampling, and Simpson's paradox biases. Historical bias refers to already-existing bias from socio-technical issues. Representation bias can be created due to the sampling phase. Measurement bias happens during selecting, utilizing, and measuring specific features. Population bias happens when the dataset distribution differs from the real-world population. Sampling bias also could be due to error while creating random subgroups. Simpson's paradox means bias from the distinction in the behavior of population subgroups in the aggregation phase [43].

Model building and evaluation bias include evaluation, aggregation, popularity, algorithmic, omitted variable, demographic, and temporal biases. Evaluation bias can arise during the model evaluation phase. Aggregation bias can happen due to the wrong assumptions about the effects of population on the model's results. Popularity occurs because of more popular items gaining more interactions [42,44]. Algorithmic bias can happen due to technical issues inside the used algorithm. Omitted variable bias takes place because of not choosing one or more essential variables for the model. Demographic bias happens due to differences in user demographic groups (e.g., age and gender) being treated differently [30,45]. Temporal bias stems from behavior and population differences with the passage of time [30,46].

Deployment and user interaction biases include behavioral, content production, linking, presentation, social, emergent, observer, interaction, and ranking biases. Behavioral bias can occur due to the dissimilarity of users' behavior in the dataset. Content production bias exists due to differences in the users' generated contents including structural, lexical, semantic, and syntactic varieties. Linking bias arises when network attributes from user activities do not truly represent the user behavior. Presentation bias happens during the presentation of information. Social bias also happens when preferences deliberately are given to certain groups and affect their judgments [30,47]. Emergent bias arises because of the difference in the real users' behavior and users' behavior in the dataset. Observer bias could happen when researchers' expectations are unintentionally injected into the research data. Interaction bias can be created due to the difference in the means of users' interaction with a system. Finally, ranking bias occurs when top-ranked results are more exposed [30].

                                                                                                                              

Due to the impact the biases have on the model's decision, it is important to consider all types of biases; however, the most recent publications mainly aimed to solve exposure bias, popularity bias, unfairness, and the bias loop effect. One of the most important challenges of the previous approaches is the trade-off between the model's performance and bias mitigation, which is believed to be based on the chosen scenario [2]. The definition of fairness also may depend on the domain, but this issue has recently drawn much attention [48].

In the recommendation area, a model that simulates multiple rounds of a bias feedback loop in a social network was proposed in [39] in order to analyze the consequence of this feedback loop in the long run. This model uses different control parameters including the level of homophily in the network, the relative size of the groups, the choice among many new link recommenders, and the choice between three various stochastic use behavior models, which decide whether each recommendation would be accepted or not. The results of this experimental study showed that a minority group with a high level of homophily can receive an excessive advantage in exposure from all link recommenders. On the other hand, if the group is heterophilic, it becomes under-exposed. Furthermore, the level of homophily in the minority group can influence the disparate exposure speed, and the relative size of the minority can magnify the effect. Both minority and majority classes based on their level of homophily can experience the "rich-get-richer" effect.

In [1], the authors worked on Conversational Recommender Systems (CRSs) and systematically investigated the popularity bias issue in state-of-the-art CRSs from various perspectives including exposure rate, success rate, and conversational utility. This article proposed a suite of popularity bias metrics that are specifically designed for CRSs. The work presented in [49] also focused on popularity bias and the long-tail problem in RSs. In addition, this paper introduced useful metrics for measuring the long-tail phenomenon on items. To complete the previously mentioned issue, Reference [30] continued this analysis further to measure algorithmic bias and fairness in a rating-based recommender system. This work considered various types of biases and fairness. Besides, this work proposed fairness metrics by analyzing two domains.

In some works including [50,51], the sensitive attribute gender was taken into consideration. Unbiased Gender Recommendation (UGRec) was introduced in [50] in order to balance performance among males and females. Aiming at seizing the users' preferences, an information aggregation component was designed to learn the representation of users and items from the user–item graph. To improve representation, a multihop mechanism was proposed by the aggregation of users' higher-order neighbors. An end-to-end training framework with adversarial learning was also used to avoid an impact on the accuracy. This framework is capable of removing gender-specific features and maintaining common features. An exploratory analysis of gender bias and discrimination in music RSs was conducted in [51]. The main aim of this work was to investigate which CF approach enhances or reduces artist gender bias. To reach this, the Preference Ratio (PR) and Bias Disparity (BD) metrics were used to measure the results. The results showed that the CF RS can amplify the gender bias problem in a real-world LastFM dataset.

Other work proposed by [52] also focused on gender bias in RSs for two book rating datasets, Amazon and book-crossing. In this research, a model-agnostic bias mitigation approach was introduced with respect to the accuracy of the system. Two recommender system approaches were used from the K-nearest neighbors' family. The results showed a significant decrease in the bias with little impact on the accuracy of the models.

### 2.4. Bias in GNN-Based RSs

Specific sensitive attributes that reinforce an already existing bias in the network of GNN-based RSs have drawn attention toward measuring fairness in supervised methods. The metrics used for this purpose make the proportion of sensitive attribute values in a protected group classified as positive to be the same as the unprotected group [14,53].

The behavior of user–item interaction does not explicitly include any sensitive information from users, but due to the high correlation between users and their attributes, directly applying modern user and item representation learning can result in the leakage of the users' sensitive information [14]. Furthermore, considering the graph-based nature of RSs, users do not have independence, and they are implicitly correlated with other users who share similar behavior, which can result in vital problems in previous models and the basics of CF recommendations [14]. In addition, current GNN algorithms are suffering from societal bias in data, which limits the generalization power of the models [15]. In the graph structure, nodes of similar sensitive attributes are prone to be connected, and this nature can result in critical bias in decision-making due to differences between representations from nodes of similar sensitive information and other nodes of other sensitive features [15]. Some approaches also consider graph embedding methods used in Online Social Networks (OSNs). Graph embedding methods are one of the best tools for data mining, which connect each user with a lower-dimensional vector that contains structural information within the network. This information can include the user's neighborhood, popularity, etc. [53]. According to previous research, OSNs suffer from discrimination, favoring the majority group, which is against anti-discrimination laws [53].

The work presented in [54] focused on calibrating the long-tail issue in session-based recommendations, which can be divided into Recurrent Neural Network-based (RNN) models and GNN-based models. This work used different metrics to evaluate the models (e.g., MRR and recall) and measured popularity bias including coverage and tail coverage. Besides, a calibration module was proposed that uses the session representation to predict the ratio of items from the tail in the recommendation list. A curriculum training strategy with two stages also was used to enhance the accuracy of predictions in the calibration module.

In [55], an investigation into graph-based Collaborative Filtering (CF) approaches for RSs was performed. In this work, two-fold performances for accuracy and novelty for currently used graph-based CF methods were taken into consideration. The results indicated that symmetric neighborhood aggregation in most of the graph-based CF models amplifies the popularity bias in RSs. In addition, this amplification can be expanded by the increase in the depth of graph propagation.

Considering works on the bias and fairness problem in GNN-based RSs, they are very limited. Most of the research works in this area focus on sensitive information in RSs. The work in [14] focused on eliminating sensitive information in representation learning, to achieve fair representation learning for a fair recommendation. To address this problem, a model-agnostic graph-based perspective for fairness-aware representation learning was introduced. The proposed model uses user and item embeddings from any recommendation models as the input and defines a sensitive feature set. In addition, the proposed model works as a filter to obscure any sensitive information in the defined set without damaging the accuracy of the recommendation. In this structure, every user can be considered an ego-centric graph structure, which helps the filters work under a graph-based adversarial training process. The discriminators were designed to make predictions of the attribute of concern, and the training of the filters was addressed to the removal of any sensitive information that can leak from the user-centric graph structure. This model was examined through two real-world datasets and showed high performance.

The aim of [15] was to overcome two major challenges in the fairness of GGN-based RSs with limited sensitive information: first, how to eradicate discrimination by fixing the insufficiency of sensitive attributes; second, the confirmation of the fairness in the GNN classifier in the RS. To tackle the mentioned issues, a new approach called FairGNN was introduced for fair node classification. To predict some of the sensitive attributes with noise, which can lead to a fair classification, an estimator for a sensitive attribute in the GNN was used in FairGNN, which can work in an atmosphere including an adversary on different datasets. The results of the experiments on real-world datasets showed that the proposed model can work effectively with respect to fairness and classification performance.

Another work [53] focused on quantifying and tackling fairness problems in graph embedding methods by using the node2vec approach for GNN-based RSs. To address this problem, this article provided a new study method for the algorithmic fairness of node2vec. In addition, the statistical parity method (which uses sensitive attributes of pairs of users to measure fairness for groups) was extended and the novel idea of the Equality of Representation was proposed to calculate fairness in a friendship RS. The node2vec approach then was applied to the real-world OSN dataset to discover biases in the recommendations caused by unfair graph embeddings. Finally, as an extension of node2vec, a new fairness-aware graph embedding algorithm called Fairwalk was introduced.

The exposure bias problem in GNN-based RSs was addressed in [56]. In this paper, a neighbor aggregation via an inverse propensity approach was proposed. The mentioned approach balances the biased local structure of each target node by obtaining the user–item propensity score for each interaction in the graph, and then, the inverse propensity score with Laplacian normalization is used as the edge weight for the process of neighbor aggregation. This leads to highlighting the less-popular neighbors in an embedding. The results showed that the debiasing method works successfully, hence increasing the performance of the model.

Considering the mentioned challenges in RSs, bias amplification is one of the most important subjects that needs to be taken into consideration. Moreover, based on the GNN structure, the accuracy of the models can be enhanced, but the bias problem can be even worse. This problem is clearly against mentioned guidelines and needs further investigation. On the other hand, sensitive attributes such as gender in GNN-based RSs are very important, and fairness toward these attributes needs to be taken into consideration.

## 3. Experimental Study

This research aimed to study the behavior of GNN-based methods against different types of biases affecting recommender systems. The main focus of this experiment was on popularity and gender bias. The results were used to determine whether this recommendation approach in most cases acquires better results in performance, but amplifies the bias. To reach this, we implemented three different types of approaches for the RS on two real-world datasets (MovieLens and LastFM): Collaborative Filtering (CF), Matrix Factorization (MF), and GNN-based approaches. We used different methods for each approach: Deep Matrix Factorization (DMF), Item-based K-Nearest Neighbor (ItemKNN), Neural Collaborative Filtering (NeuMF), Neural Collaborative Filtering Model with Interaction-based Neighborhood (NNCF), Neural Graph Collaborative Filtering (NGCF), a Light version of Graph Convolution Network for recommendation (LightGCN), and Self-supervised Graph Learning (SGL). We compared the results based on different metrics for both performance and bias.

### 3.1. Benchmark Datasets

In this research, two real-world datasets were used for implementing the RS. These two datasets suffer from the mentioned long-tail effect on items, which denotes the existence of popularity bias in the data, which makes them a good choice for bias investigation. In the section below, we give a description of these two datasets.

### 3.1.1. MovieLens 100K [57]

MovieLens is one of the widely used datasets in the field of RSs and especially in bias investigation. This dataset is collected gradually from the MovieLens website, a non-commercial web-based movie recommender system, and randomly selected. The dataset includes ratings of users on movies on a star scale in the interval [1, 5]. Among the different datasets in different sizes provided on this website, we used in this project the ml-100 K dataset, which includes 100-thousand records of ratings. MovieLens, moreover, consists of three different files, which contain, respectively, the information related to the users, the items, and the ratings given to the items by the users. The sensitive features are located in the users' dataset, where, based on capAI guidance [58], "Gender" and "Age" are detected as sensitive features. This dataset also suffers from popularity bias on items. This means popular movies received more ratings in comparison to the other movies. The mentioned reasons make this dataset a very good fit for this investigation into the bias problem. The MovieLens dataset information can be seen in Table 1.

**Table 1.** MovieLens dataset information.

| Features | Description | Data Type | Count | Mean | Std |
|----------|-------------|-----------|-------|------|-----|
| Age | Age of users | int | 100 K | 32.96 | 11.56 |
| Rating | Rating on movies provided by users | float | 100 K | 3.52 | 1.12 |
| User id | IDs of the users | int | 100 K | - | - |
| Movie id | IDs of the movies | int | 100 K | - | - |
| Gender | Gender of the user | String | 100 K | - | - |
| Occupation | Users' job | String | 100 K | - | - |
| Movie title | The title of rated movies | String | 100 K | - | - |

In this research, Exploratory Data Analysis (EDA) is provided in order to have a better understanding of both datasets. The long-tail effect can be seen in Figure 1.

Figure 1 shows that the distribution of ratings on items in the MovieLens dataset is heavily focused on popular items. This means this dataset contains popularity bias toward items. The number of items in this dataset is 1682, 87.3% of which are in the long-tail.

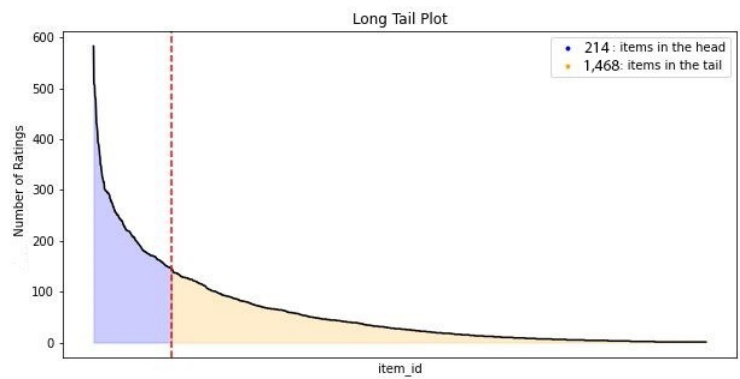

**Figure 1.** MovieLens long-tail plot.

Figure 2 indicates the number of ratings for each rating segment, which is defined based on the ratings of the users on items from 1 to 5. This plot shows the distribution on average of the ratings ranging from 1 to 5.

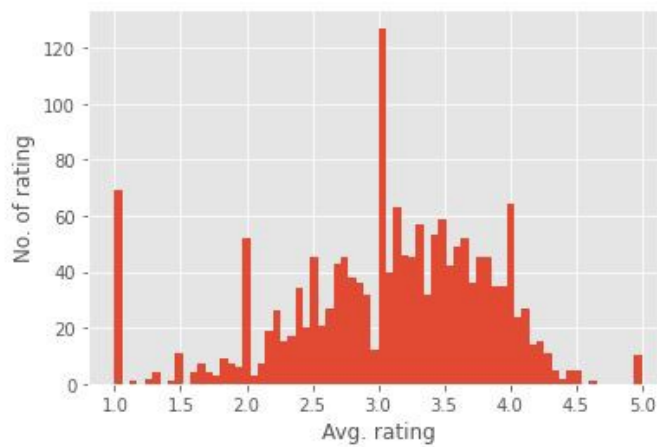

**Figure 2.** MovieLens average ratings plot.

Figure 3 illustrates the distribution of age and gender, which shows that a majority of the users are young individuals and the number of rated items by men is significantly more than by women.

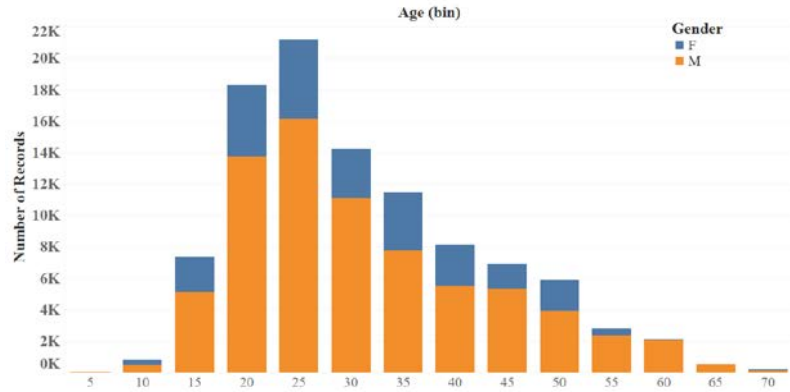

**Figure 3.** MovieLens age/gender distribution.

Figure 4 shows that the number of male users is considerably higher than that of women.

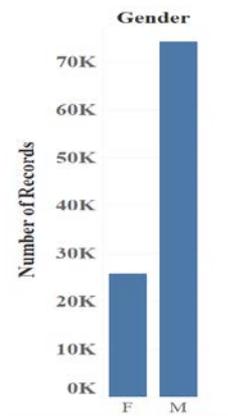

**Figure 4.** MovieLens gender distribution.

Figure 5 shows the number of rated items based on the occupation of the users. We can see that the predominant occupation of the users in the dataset is student.

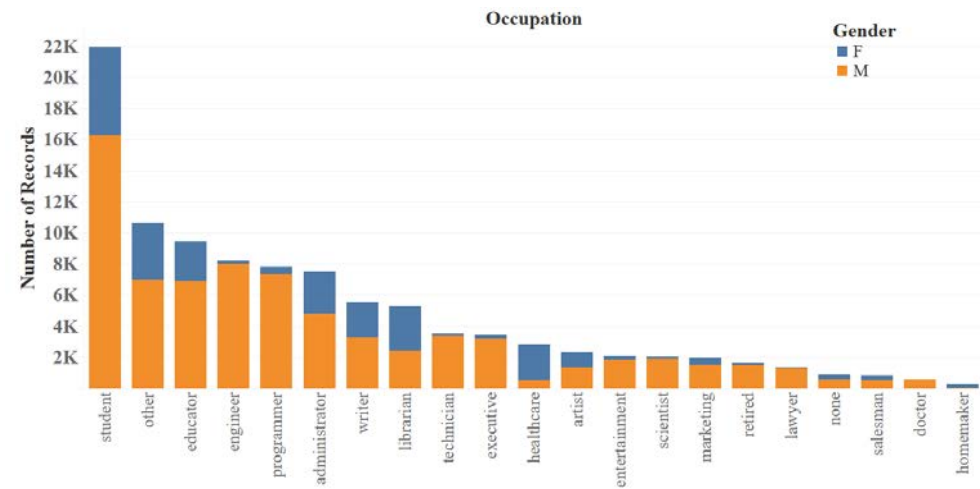

**Figure 5.** MovieLens occupation distribution.

### 3.1.2. LastFM [59]

The LastFM dataset also is widely used in RSs, especially when it comes to dealing with the popularity bias problem. This dataset is one of the largest datasets, which includes user and artist information from all over the world. This dataset shows how many times each user has listened to each artist. The LastFM dataset includes the features of users and artists and the interactions among them. According to the capAI guidance [58], gender can be considered a sensitive feature in this dataset. The analysis of this dataset shows that the ratings of popular items are significantly higher than the other items. The LastFM dataset's information and details can be seen in Table 2.

**Table 2.** LastFM dataset information.

| Features | Description | Data Type | Count | Mean | Std |
|----------|-------------|-----------|-------|------|-----|
| Weight | Listening count for each artist | float | 100 K | 745.24 | 3751.32 |
| User id | IDs of the users | int | 100 K | - | - |
| Item id | IDs of the artists | int | 100 K | - | - |
| Gender | Age of users | String | 100 k | - | - |
| Country | Users' country | String | 100 K | - | - |
| Name | Names of the artists | String | 100 K | - | - |

In the following section, we provide an EDA on the LastFM dataset.

Figure 6 illustrates the distribution of the ratings of artists in the LastFM dataset. The figure shows a high popularity bias due to the fact that the ratings are concentrated on a small number of items. Specifically, 90.3% out of the 41,269 items in the dataset are in the long-tail.

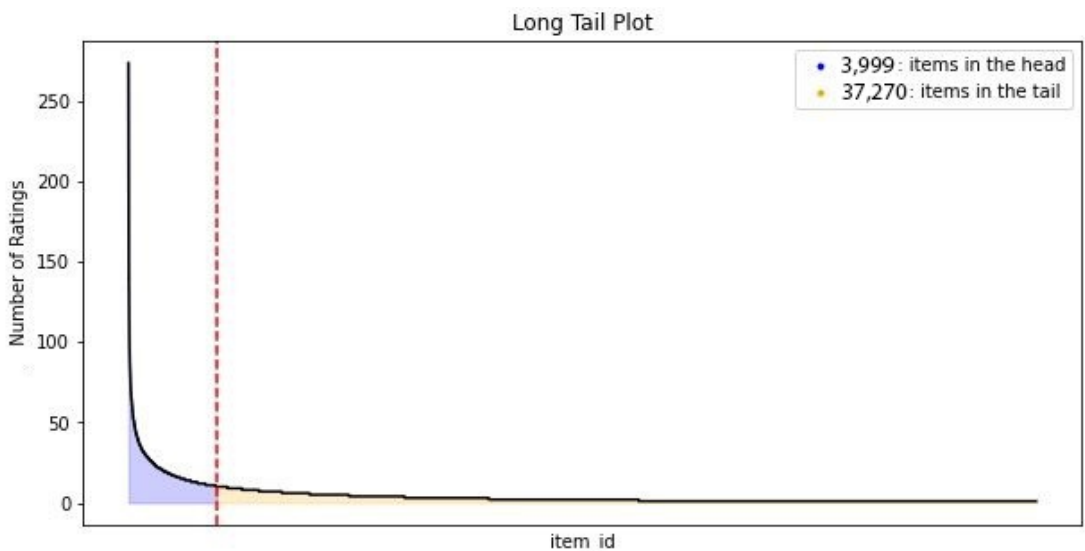

**Figure 6.** LastFM long-tail distribution.

Figure 7 shows the gender distribution for the LastFM dataset. As in the MovieLens dataset, the number of male users is considerably higher than the female users.

Figure 8 illustrates the top ten most popular artists regarding the number of records.

Figure 9 shows the distribution of records throughout the world map. It can be seen that most of the records belong to the U.S.

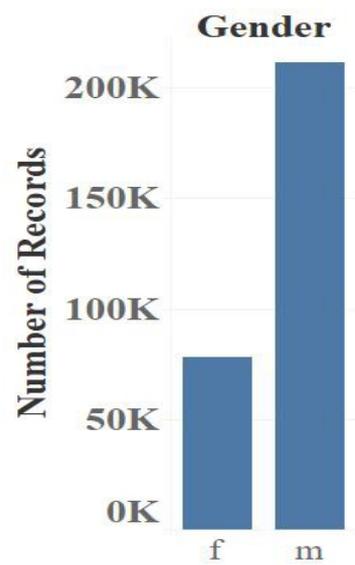

**Figure 7.** LastFM gender distribution.

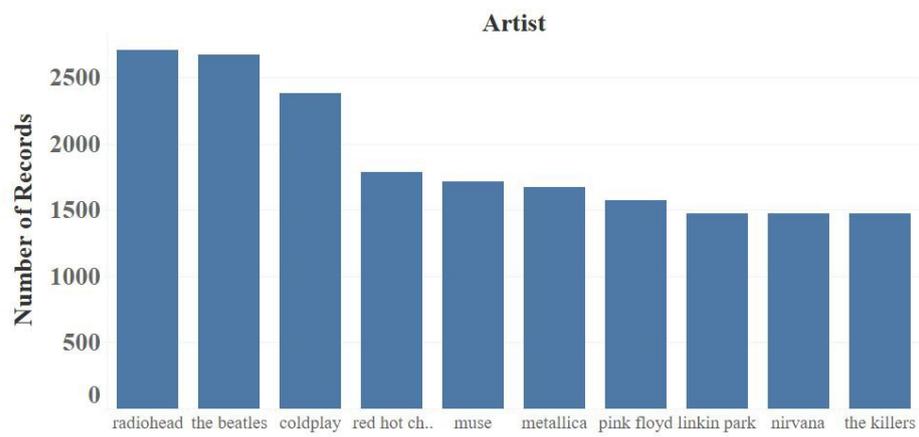

**Figure 8.** LastFM top ten artists.

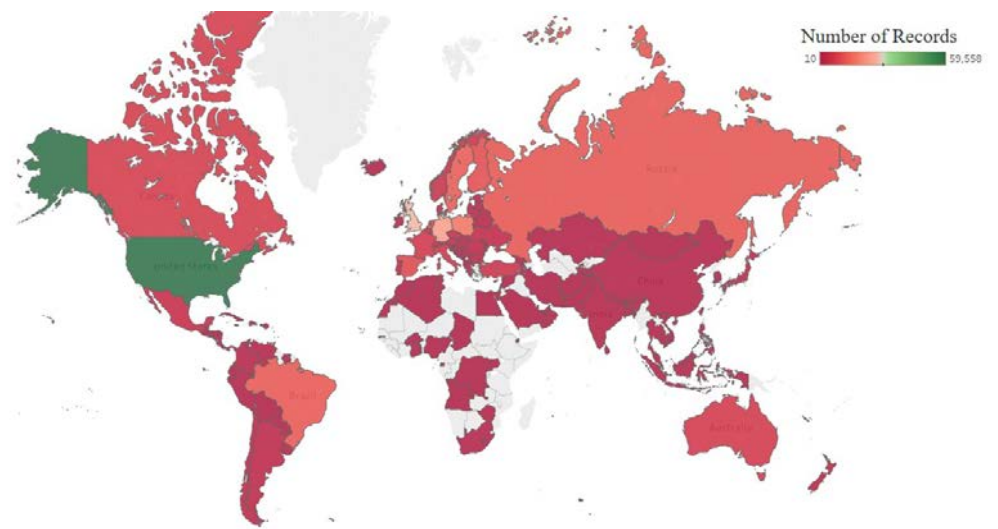

**Figure 9.** LastFM distribution of the records on the map.

### 3.2. Recommendation Methods

In this research study, we used three recommendation approaches for RSs: Collaborative Filtering (CF), Matrix Factorization (MF), and GNN-based recommendation methods. We also implemented different methods for each approach to gather a wide range of results, which allowed us to reach more sound conclusions. Implementing different methods can help us to analyze the bias problem further; hence, different models should be compared to investigate bias amplification in detail. Below is the description of these approaches and chosen models:

1. Collaborative Filtering (CF):

    Collaborative (or social) filtering techniques are based on user preferences and use information about ratings given by users to items to compute user or item similarity. If two users rate items in a similar way, it is more likely that they rate the new items likewise. Therefore, the target user is one who recommended items well rated by users with the same taste. Another CF strategy is to recommend items similar to those that the user has consumed or has rated positively; such similarity between items is calculated from the ratings received by users. In CF approaches, user ratings on items are collected in order to create a user–item rating matrix. This matrix is used to find similarities between users/items. CF approaches can tackle some of the problems of content-based approaches, in which items similar to those consumed or well rated by the user are also recommended, but this similarity is computed from item features. A drawback of this approach, which is avoided by CF, is the unavailability of item features or the difficulty to obtain them, since recommendations in CF are made using only the feedback of other users. Besides, the quality of these techniques is higher because they are based on items evaluated by users, instead of relying on content, whose quality can be low. CF approaches unlike content-based systems can recommend items with various content, that is not similar to those previously consumed by the user, as long as other users have already shown interest in these different items. CF techniques use different approaches including:

    - User-based: these systems assess the preference of a target user for an item using the ratings given to this item by his/her neighbors, which are users that have a similar rating behavior [4].
    - Item-based: these approaches anticipate the rating of a user for an item considering the ratings given by this user to similar items. In such approaches, two items are similar if they have received similar ratings from several users in the system. This is different from content-based methods, which base the similarities of items on their characteristics or attributes. This approach is more convenient in common commercial recommender systems, where the number of users is much higher than the number of items in the catalog. Usually, item-based approaches are more reliable, require less computation time, and do not need to be updated as frequently [4].

    Figure 10 shows the differences between user-based and item-based approaches. The CF approaches used are as follows:

    - ItemKNN: This method is an item-based approach that computes the similarity between the items based on the ratings that users give to them. The main motivation behind this method is that customers are more prone to purchase items that are compatible with their previous purchases. Historical purchase information in the user–item matrix can lead to recognizing sets of similar items and using them to create the top-K recommendations. This algorithm in a high-level view includes two major elements. The first component creates a model seizing the relations among various items. The second component applies the mentioned calculated model to acquire top-K recommendations for a user. This method also shows a high performance in comparison to other similar CF approaches [60–62].

- Neural Collaborative Filtering model with interaction-based Neighborhood (NNCF): This model utilizes deep learning for modeling complicated interactions between users and items. This novel model also uses neighborhood information to complete user–item interaction data, hence improving the model's performance. NNCF models can overcome traditional algorithmic issues such as simple linear factorization, which may not completely support complex interaction among users and items. This method can also provide user/item embeddings with a good quality [63–65].

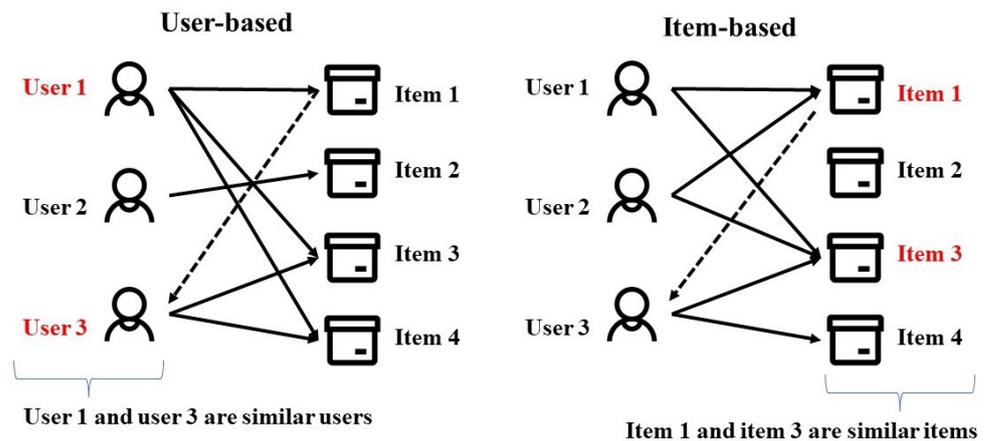

**Figure 10.** The Differences between user-based and item-based approaches.

2.  Matrix factorization:

    Matrix factorization encompasses a group of model-based techniques in which the rating matrix is transformed into two matrices of latent factors representing users and items, respectively, in an attempt to tackle the sparsity problem of the ratings matrix. This is a low-dimensional factor model, where it is assumed that the inner products between the user and item latent factors influence the preferences of the user for an item [66]. Currently, MF has become one of the most popular methods for implementing RS [67].

    The MF approaches used are as follows:

    - Deep Matrix Factorization (DMF): This method uses a neural network architecture. This method constructs the user–item matrix with explicit ratings and non-preference implicit feedback. Afterward, this matrix is used as the input for learning a common low-dimensional space for the deep structure learning architecture. This method also uses a novel loss function based on binary cross-entropy, which considers both explicit ratings and implicit feedback for enhancing optimization. DMF provides better top-K recommendations in comparison to traditional models by applying implicit feedback, thus reconstructing the users' ratings via learning hidden structures with explicit historical ratings. This method also supports two-channel structures, which can combine side information from both users and items. Some articles also indicate that this approach can outperform new recommendation algorithms with respect to accuracy and training efficiency [68–70].

    - Neural Collaborative Filtering (NeuMF): Knowing that the most important factor in CF models is the interaction between user and item features, the inner products in these methods can be replaced by a neural network architecture. Neural network-based Collaborative Filtering (NCF) is a schema that expresses and generalizes matrix factorization, which can be enhanced by using nonlinear kernels. To achieve this, a multi-layer perceptron can be used in order to learn the user–item interaction function [71]. The good capacity and nonlinearity of deep neural networks are the main reasons for their good performance. Furthermore,

the general NCF used in NeuMF can provide us with the opportunity for using the combination of various models [67].

3. GNN-based:

One of the fastest-growing technologies that had great capability in recent years is Graph Learning (GL) [14]. This approach relates to machine learning applied to graph structure data. Using these advantages to learn relational data, Graph Learning-based Recommender Systems (GLRSs) have been proposed [6]. In reality, the majority of objects are explicitly or implicitly connected with each other. These relations can be shown by graphs. In RSs where the objects can be considered users, items, attributes, and context, this characteristic is even clearer.

These objects are strongly connected with each other and affect each other via different relations. The quality of RSs can be remarkably increased by using graph techniques. Graph learning has a great ability to learn complex relations, as well as a high potential in obtaining knowledge enclosed in a variety of graphs [72].

There are different types of entities in RSs including users, items, and attributes, which maintain different types of relationships with each other and can therefore be represented by graphs of diverse types. It is well known that the three main objects used in recommender models are user, item, and user–item interaction, although other information concerning users and/or items may also be used. On this basis, data used in RSs can be classified into two broad categories: user–item interaction data (clicks, purchases, or ratings made by the users on the items) and side information data (user and item attributes). In addition, interaction data can be classified into two categories depending on whether the interactions are sequential or general. [72]. Each class also is divided into various sub-classes, as can be seen in Table 3.

**Table 3.** A summary of data representation in RSs and representing graphs [72].

| Data Class | Data Subclass | Representing Graph |
|---|---|---|
| General interaction | Explicit interaction, implicit interaction | Weighted bipartite graph, unweighted bipartite graph |
| Sequential interaction | Single-type interactions multi-type interactions | Directed homogeneous graph, directed heterogeneous graph |
| Side information | Attribute information, social information, external knowledge | Heterogeneous graph, homogeneous graph tree, or heterogeneous graph |

Each input in the user–item matrix is information about the type of interaction that happened between them. The interaction data can be divided into categories based on their types: explicit and implicit. Explicit interaction happens when a user is asked to provide an opinion on an item (e.g., users' ratings on items). Implicit interaction is the one that is concluded from the user's action (e.g., click, view) [72,73].

The GNN methods used are the following:

- LightGCN: This model is a simple version of a Graph Convolution Network (GCN), which includes the most important components of GCNs for recommendation tasks. LightGCN linearly propagates the user and item embeddings on the user–item interaction graph. Afterward, this model uses the weighted sum of the embeddings learned at all layers as the final embedding [74]. The symmetric normalization in LightGCN is the same as the standard GCN, which controls the increase in the size of embeddings with graph convolution operations. This method also showed great performance in comparison to conventional approaches [75,76].

- Neural Graph Collaborative Filtering (NGCF): This model is another chosen method for this investigation. This model introduces a graph structure into user–item interactions. This method benefits from the user–item graph structure by generating embeddings on it, which results in high-order connectivity in the user–item graph. The collaborative signal is pumped into the embedding process in an explicit way [72]. This method, moreover, uses multiple embedding propagation layers, with concatenated outputs to create the final prediction for the recommendation task. NGCF also shows great performance concerning model optimization [77].

- Self-supervised Graph Learning (SGL): This method is an improved version of GCN models with respect to accuracy and robustness. These models also perform better while working with interactions with noise. This method uses an enhanced classical supervised task of recommendation with a supporting self-supervised task, which reinforces node representation learning via self-discrimination. This structure generates different views of a node, which maximizes the agreement between various views of the same node compared to that of the other nodes. Three operators are also devised in order to generate the mentioned views—node dropout, edge dropout, and random walk—which change the graph structure in different aspects. The SGL method has also shown great performance in RS tasks, which makes it a suitable choice for this experiment [78–80].

### 3.3. Evaluation Metrics

In order to evaluate the models implemented with the previously described methods from the point of view of the reliability of the recommendations, as well as from the perspective of sensitivity to biases, we used both performance metrics and metrics for measuring bias amplification. Considering there is a set of $n$ items to be ranked and given an item set $I$ and a user set $U$, $\hat{R}(u)$ is used to represent a ranked list of items that a model produces and $R(u)$ represents a ground-truth set of items that user $u$ has interacted with. For top-K recommendations, only top-ranked items need to be considered. Therefore, in top-K evaluation scenarios, we truncated the recommendation list with a length $K$.

In order to have a better understanding of the metrics, a notation table is provided below in Table 4.

**Table 4.** Table of notations.

| Notation | Definition |
|----------|------------|
| $U$ | A set of users |
| $V$ | A set of items |
| $u$ | A user |
| $v$ | An item |
| $R(u)$ | A ground-truth set of items that user $u$ interacted with |
| $\hat{R}(u)$ | A ranked list of items that a model produces |
| $K$ | The length of the recommendation list |
| $M(x)$ | Algorithmic mechanism for the RS with input $x$ and output $y$ |
| $\theta$ | Distribution, which generates $x$ |
| $\Theta$ | A set of distributions of $\theta$, which generate each instance $x$ |

Table 5 shows the main metrics for assessing the reliability of the recommendations. As in most current recommender systems, the evaluation was performed on top-K item recommendation lists, where K represents the size of the list.

Table 6 shows the metrics used in this study to evaluate the sensitivity of the models to the most relevant types of biases in recommender systems. In this research, we mainly focused on bias amplification related to the popularity and diversity of the recommended items. To reach this, three different metrics (average popularity, Gini Index, and item coverage) were used, as can be seen in Table 6. In addition, we evaluated the gender bias by means of the Differential Fairness (DF) metric, also described in Table 6. The objective of this metric is to evaluate whether the recommendation algorithms behave the same for all values of the protected attributes or produce a certain degree of bias in the output for some of the values. In this study, we considered the values "male" and "female" of the protected attribute "gender". This implies that the probabilities of the predicted item scores should be similar for both values of this attribute. DF specifically focuses on both intersectionality [81,82], which involves fairness for each of the protected attributes individually, and behavior toward minorities, which is related to the previously mentioned anti-discrimination laws.

**Table 5.** Metrics for performance evaluation.

| Metric Name | Description |
| --- | --- |
| MRR | Computes the reciprocal rank of the first relevant item found by an algorithm. Considers $Rank_u^*$ to be the rank position of the first relevant item found by an algorithm for a user u. <br><br> $$MRR@K = \frac{1}{|U|} \sum_{u \in U} \frac{1}{Rank_u^*}$$ |
| NDCG | Is a measure of ranking quality, where positions are discounted logarithmically. It accounts for the position of the hit by assigning higher scores to hits at top ranks. $\delta(0)$ is an indicator function. <br><br> $$NDCG@K = \frac{1}{|U|} \sum_{u \in U} \frac{1}{\sum_{i=1}^{min(|R(u)|,K)} \frac{1}{log_2(i+1)}} \sum_{i=1}^{K} \delta(i \in R(u)) \frac{1}{log_2(i+1)}$$ |
| Precision | Is a measure for computing the fraction of relevant items out of all the recommended items. Its final value is the average of the metric values for each user. $|\hat{R}(u)|$ represents the item count of $\hat{R}(u)$. <br><br> $$Precision@K = \frac{1}{|U|} \sum_{u \in U} \frac{|\hat{R}(u) \cap R(u)|}{|R(u)|}$$ |
| Recall | Is a measure for computing the fraction of relevant items out of all relevant items. $|R(u)|$ represents the item count of $R(u)$. <br><br> $$Recall@K = \frac{1}{|U|} \sum_{u \in U} \frac{|\hat{R}(u) \cap R(u)|}{|R(u)|}$$ |
| HR (HIT) | This metric is also known as the truncated hit-ratio. It is a way of calculating how many "hits" are included in a K-sized list of ranked items. If there is at least one item that falls in the ground-truth set, we call it a hit. $\delta(0)$ is an indicator function. $\delta(b) = 1$ if $b$ is true; otherwise, it would be 0. $\varnothing$ denotes the empty set. <br><br> $$HR@K = \frac{1}{|U|} \sum_{u \in U} \delta(\hat{R}(u) \cap R(u) \ne \varnothing)$$ |

**Table 6.** Metrics for bias measurement.

| Metric Name | Description |
| --- | --- |
| Average Popularity | Computes the average popularity of recommended items. In the formula below, $\varphi(i)$ is the number of interaction on item i in the training data [83]. $$AveragePopularity@K = \frac{1}{|U|} \sum_{u \in U} \frac{\sum_{i \in R_u} \varphi(i)}{|R_u|}$$ |
| Gini Index [41] | Presents the diversity of the recommendation items. It is used to measure the inequality of a distribution. In the following formula, $P(i)$ represents the number of times that item $i$ appears in the recommended list, which is indexed in non-decreasing order [84]. $$GiniIndex@K = \frac{\sum_{i=1}^{|I|}(2i - |I| - 1)P(i)}{|I| \sum_{i=1}^{|I|} P(i)}$$ |
| Item Coverage | Computes the coverage of recommended items over all items [85] $$ItemCoverage@K = \frac{|\cup_{u \in U} \hat{R}(u)|}{|I|}$$ |
| Differential Fairness (DF) for sensitive attribute gender [81,86] | Ensures unbiased treatment for all protected groups. This metric also denotes the privacy interpretation of disparity. The mechanism $M(x)$ is $e$-differentially fair with respect to $(A, \theta)$ for all $\theta \in \Theta$ with $x \sim \theta$ and $y \in Range(M)$. For all $(s_i, s_j) \in A \times A$, where $P(s_i) > 0$, $P(s_j) > 0$. $s_i, s_j \in A$ are tuples of all protected attribute values (here, male and female). $$e^{-e} \leq \frac{P_{M,\theta}(M(x) = y | s_i, \theta)}{P_{M,\theta}(M(x) = y | s_j, \theta)} \leq e^e$$ |

## 4. Results

In this section, we present the results provided by the recommendation methods described in the previous section on the datasets MovieLens and LastFM. These results came from applying the metrics explained above. The objective was to evaluate the different recommendation methods to determine which ones present the best balance between performance and sensitivity to biases, since the improvement of the former usually leads to a worsening of the latter. The results will also serve to determine whether the hypothesis that GNN-based methods produce more reliable, but also more biased models is confirmed. This would evidence the need for further research on ways to deal with the amplification of biases in these methods.

In Tables 7 and 8, the results of the mentioned models can be seen on the Movie-Lens and LastFM datasets. Different metrics also are provided in order to have a better understanding of the performance of the models and bias amplification for each model.

Table 7 shows the results on the LastFM dataset.

**Table 7.** Results on the MovieLens dataset.

| Approach | Method | Top K | Recall | Precision | MRR | NDCG | HIT | Item Coverage | Gini Index | Average Popularity |
|----------|--------|-------|--------|-----------|-----|------|-----|---------------|------------|--------------------|
| MF | DMF | K = 5 | 0.14 | 0.22 | 0.43 | 0.26 | 0.62 | 0.18 | 0.94 | 256.29 |
| MF | DMF | K = 10 | 0.21 | 0.17 | 0.42 | 0.25 | 0.73 | 0.20 | 0.93 | 252.25 |
| MF | DMF | K = 15 | 0.29 | 0.16 | 0.45 | 0.28 | 0.83 | 0.28 | 0.90 | 219.49 |
| MF | NeuMF | K = 5 | 0.15 | 0.23 | 0.45 | 0.27 | 0.65 | 0.25 | 0.91 | 228.52 |
| MF | NeuMF | K = 10 | 0.23 | 0.18 | 0.46 | 0.27 | 0.78 | 0.36 | 0.89 | 212.41 |
| MF | NeuMF | K = 15 | 0.30 | 0.16 | 0.46 | 0.28 | 0.83 | 0.40 | 0.86 | 196.89 |
| CF | ItemKNN | K = 5 | 0.15 | 0.23 | 0.44 | 0.28 | 0.63 | 0.19 | 0.93 | 231.96 |
| CF | ItemKNN | K = 10 | 0.22 | 0.18 | 0.46 | 0.27 | 0.75 | 0.24 | 0.93 | 249.74 |
| CF | ItemKNN | K = 15 | 0.31 | 0.16 | 0.46 | 0.29 | 0.84 | 0.29 | 0.89 | 208.12 |
| CF | NNCF | K = 5 | 0.15 | 0.24 | 0.47 | 0.29 | 0.64 | 0.17 | 0.95 | 284.47 |
| CF | NNCF | K = 10 | 0.24 | 0.19 | 0.46 | 0.22 | 0.78 | 0.25 | 0.91 | 217.70 |
| CF | NNCF | K = 15 | 0.28 | 0.15 | 0.47 | 0.27 | 0.81 | 0.30 | 0.91 | 231.28 |
| GNN | NGCF | K = 5 | 0.15 | 0.24 | 0.48 | 0.29 | 0.66 | 0.15 | 0.95 | 277.85 |
| GNN | NGCF | K = 10 | 0.25 | 0.20 | 0.49 | 0.30 | 0.77 | 0.25 | 0.93 | 255.49 |
| GNN | NGCF | K = 15 | 0.32 | 0.17 | 0.49 | 0.31 | 0.86 | 0.32 | 0.89 | 219.13 |
| GNN | LightGCN | K = 5 | 0.11 | 0.17 | 0.36 | 0.21 | 0.55 | 0.05 | 0.98 | 245.13 |
| GNN | LightGCN | K = 10 | 0.18 | 0.14 | 0.37 | 0.21 | 0.67 | 0.07 | 0.97 | 312.47 |
| GNN | LightGCN | K = 15 | 0.23 | 0.12 | 0.38 | 0.21 | 0.76 | 0.10 | 0.96 | 292.8 |
| GNN | SGL | K = 5 | 0.15 | 0.25 | 0.47 | 0.29 | 0.66 | 0.24 | 0.91 | 229.24 |
| GNN | SGL | K = 10 | 0.25 | 0.20 | 0.49 | 0.29 | 0.80 | 0.31 | 0.89 | 209.39 |
| GNN | SGL | K = 15 | 0.31 | 0.17 | 0.49 | 0.30 | 0.85 | 0.34 | 0.88 | 200.63 |

Table 8 shows the results on the LastFM dataset.

**Table 8.** Results on the LastFM dataset.

| Approach | Method | Top K | Recall | Precision | MRR | NDCG | HIT | Item Coverage | Gini Index | Average Popularity |
|----------|--------|-------|--------|-----------|-----|------|-----|---------------|------------|--------------------|
| MF | DMF | K = 5 | 0.05 | 0.05 | 0.11 | 0.05 | 0.20 | 0.01 | 0.99 | 377.81 |
| MF | DMF | K = 10 | 0.07 | 0.03 | 0.12 | 0.06 | 0.25 | 0.02 | 0.99 | 341.85 |
| MF | DMF | K = 15 | 0.08 | 0.02 | 0.12 | 0.07 | 0.30 | 0.02 | 0.99 | 309.64 |
| MF | NeuMF | K = 5 | 0.10 | 0.10 | 0.25 | 0.12 | 0.40 | 0.05 | 0.98 | 167.49 |
| MF | NeuMF | K = 10 | 0.15 | 0.07 | 0.27 | 0.14 | 0.52 | 0.06 | 0.98 | 157.12 |
| MF | NeuMF | K = 15 | 0.20 | 0.06 | 0.27 | 0.16 | 0.60 | 0.09 | 0.98 | 140.17 |
| CF | ItemKNN | K = 5 | 0.12 | 0.11 | 0.29 | 0.14 | 0.41 | 0.12 | 0.96 | 152.64 |
| CF | ItemKNN | K = 10 | 0.16 | 0.08 | 0.30 | 0.16 | 0.50 | 0.23 | 0.93 | 131.54 |
| CF | ItemKNN | K = 15 | 0.20 | 0.06 | 0.30 | 0.18 | 0.57 | 0.31 | 0.91 | 118.00 |
| CF | NNCF | K = 5 | 0.09 | 0.07 | 0.16 | 0.09 | 0.31 | 0.04 | 0.98 | 195.14 |
| CF | NNCF | K = 10 | 0.12 | 0.06 | 0.17 | 0.10 | 0.38 | 0.05 | 0.98 | 185.23 |
| CF | NNCF | K = 15 | 0.15 | 0.05 | 0.19 | 0.12 | 0.49 | 0.06 | 0.99 | 177.06 |
| GNN | NGCF | K = 5 | 0.12 | 0.11 | 0.29 | 0.14 | 0.44 | 0.03 | 0.99 | 202.55 |
| GNN | NGCF | K = 10 | 0.18 | 0.09 | 0.32 | 0.17 | 0.59 | 0.06 | 0.98 | 155.32 |
| GNN | NGCF | K = 15 | 0.21 | 0.07 | 0.31 | 0.18 | 0.64 | 0.08 | 0.98 | 160.38 |
| GNN | LightGCN | K = 5 | 0.13 | 0.14 | 0.31 | 0.15 | 0.47 | 0.05 | 0.98 | 174.23 |
| GNN | LightGCN | K = 10 | 0.19 | 0.09 | 0.33 | 0.18 | 0.59 | 0.09 | 0.98 | 148.15 |
| GNN | LightGCN | K = 15 | 0.23 | 0.07 | 0.34 | 0.20 | 0.66 | 0.12 | 0.97 | 132.52 |
| GNN | SGL | K = 5 | 0.13 | 0.13 | 0.33 | 0.15 | 0.48 | 0.06 | 0.98 | 142.71 |
| GNN | SGL | K = 10 | 0.20 | 0.10 | 0.35 | 0.19 | 0.62 | 0.10 | 0.97 | 114.49 |
| GNN | SGL | K = 15 | 0.24 | 0.07 | 0.35 | 0.21 | 0.69 | 0.14 | 0.96 | 103.11 |

Figure 11 shows the recall values for three different sizes of the top-K lists of the implemented models on the two datasets. According to these results, two of the three GNN-based methods, NGCF and SGL, showed better performance for both datasets regarding the recall metric, while the third one, LightGCN, showed a different behavior in each dataset.

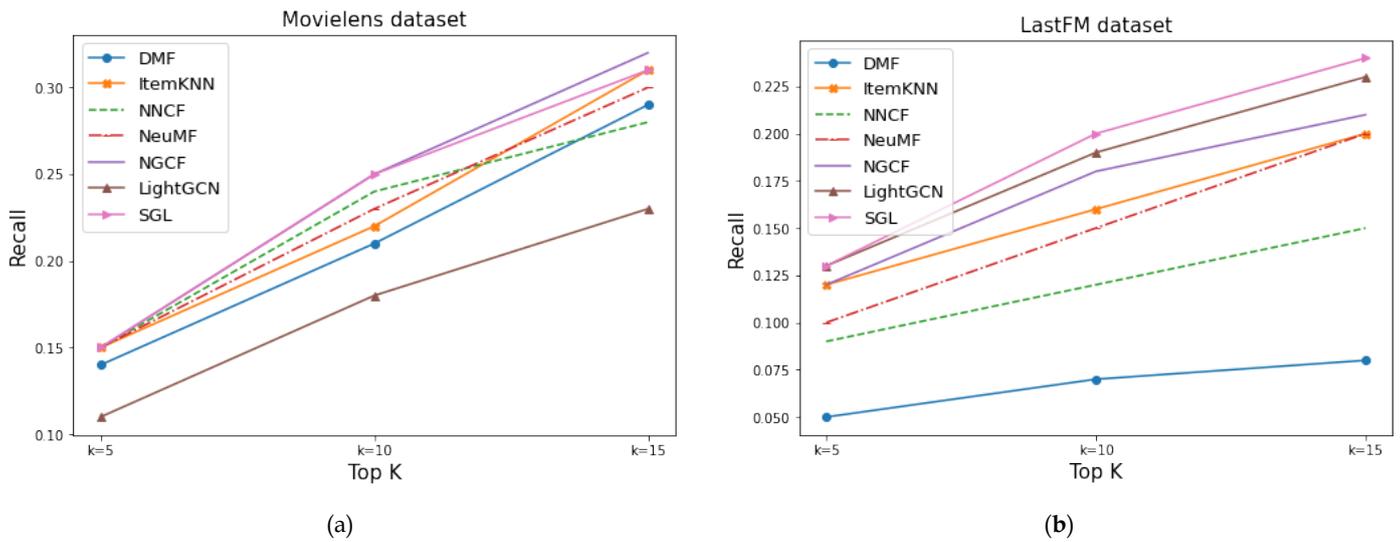

(a)                                                                (b)

**Figure 11.** Results of recall for Movielens and LastFM. (**a**) MovieLens recall. (**b**) LastFM recall.

Regarding the precision of the given models, shown in Figure 12, the results were similar. From this plot, it can be seen that NGCF and SGL provided better precision for both datasets, but LightGCN could also be a good choice for the LastFM dataset.

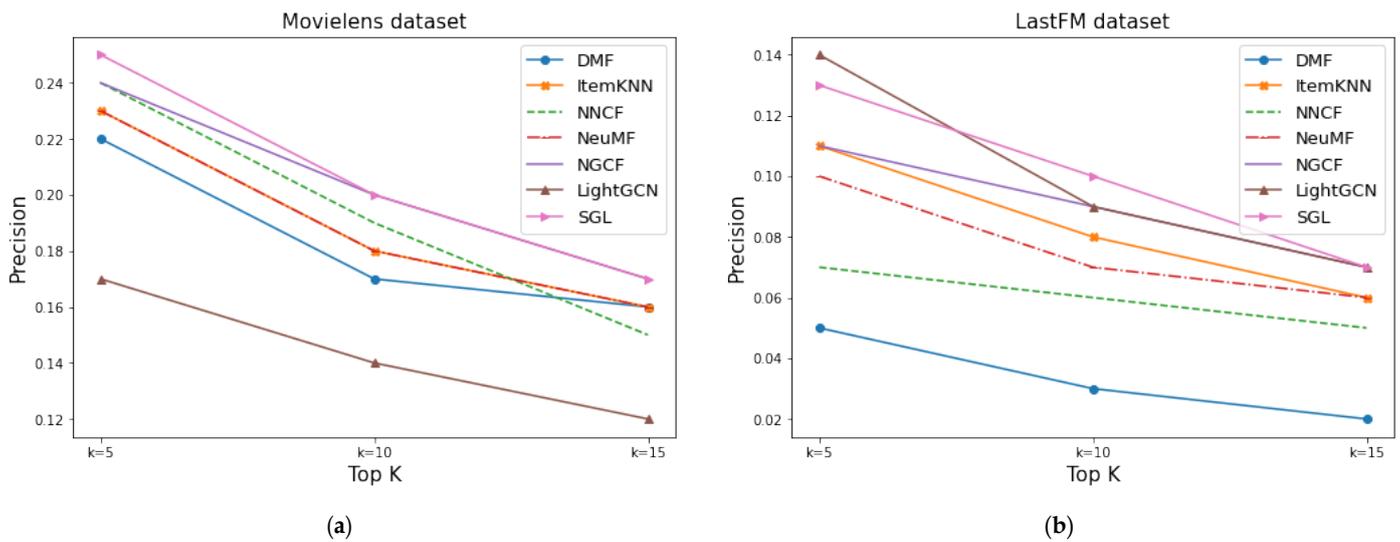

(a)                                                                (b)

**Figure 12.** Results of precision for Movielens and LastFM. (**a**) MovieLens precision. (**b**) LastFM precision.

Figure 13 illustrates the MRR metric. The results showed, just like the previous metrics, that the NGCF and SGL models provided a higher MRR. LightGCN also showed a reasonable value for the LastFM dataset.

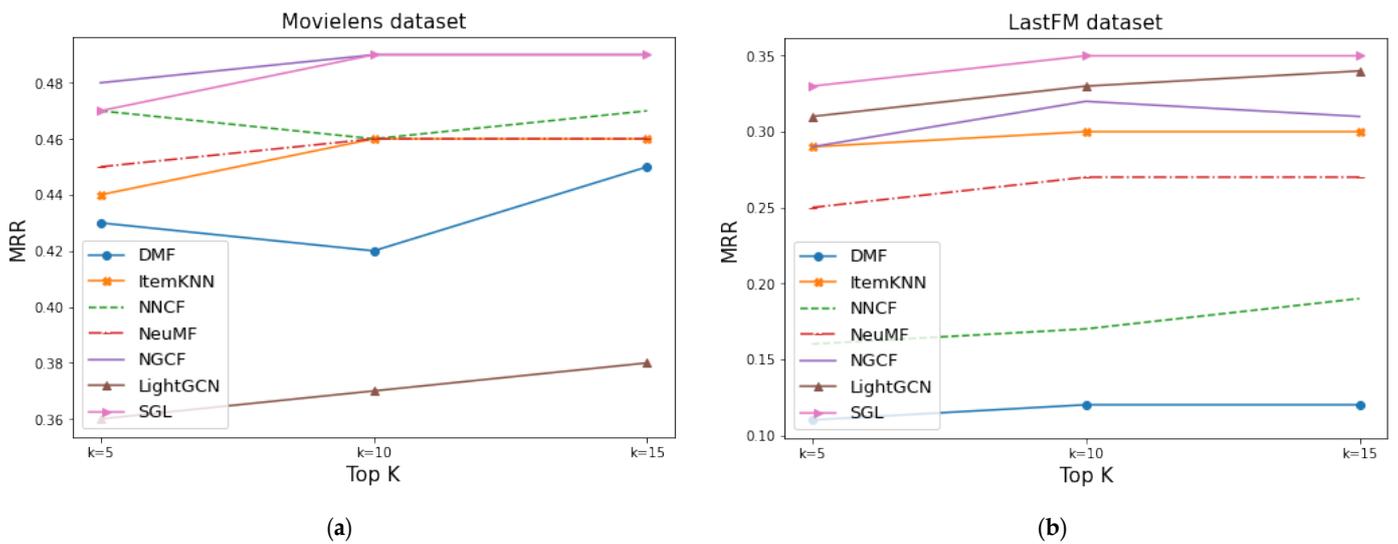

(**a**)                                                                                    (**b**)

**Figure 13.** Results of MRR for Movielens and LastFM. (**a**) MovieLens MRR. (**b**) LastFM MRR.

Figure 14 shows the HIT measure. From this metric's view, most of the implemented models showed similar results for MovieLens except LightGCN, which performed considerably poorly, although SGL and NGCF obtained the highest values. On the other hand, this metric showed that SGL, LightGCN, and NGCF, the three GNN-based methods, provided a higher HIT for the LastFM dataset.

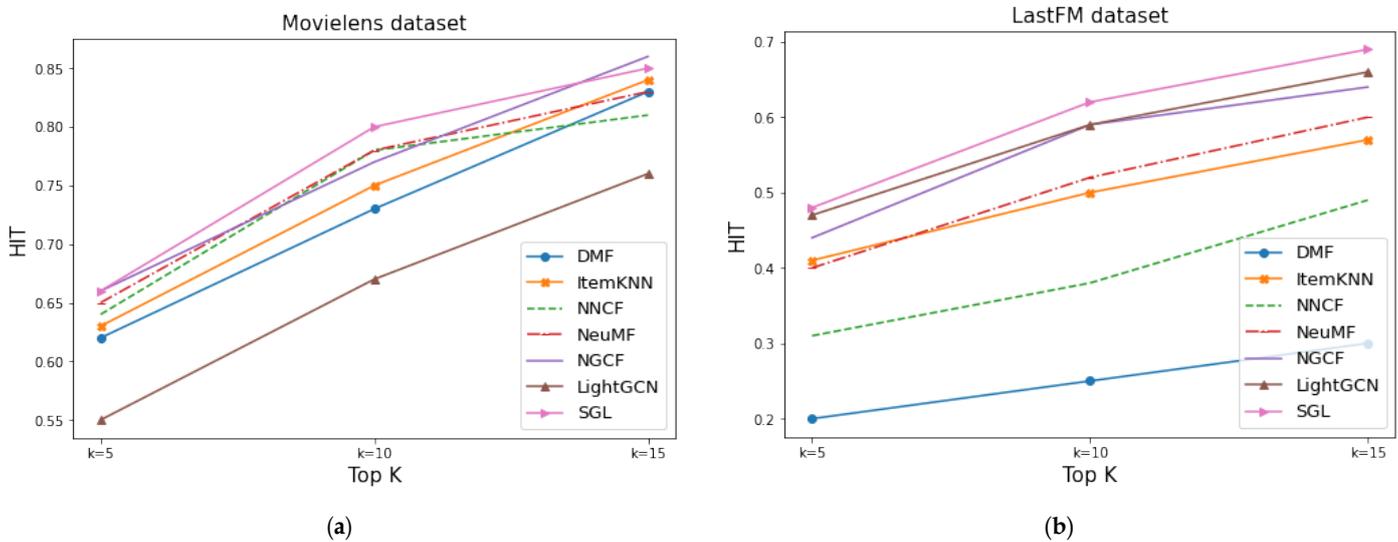

(**a**)                                                                                    (**b**)

**Figure 14.** Results of HIT for Movielens and LastFM. (**a**) MovieLens HIT. (**b**) LastFM HIT.

Figure 15 shows the NDCG of the implemented models. We can observe in the graph that two GNN-based methods, NGCF and SGL, provided better results for Movielens than the other methods. On the LastFM dataset, the three GNN-based algorithms gave the best values of NDCG, although LightGCN performed better than NGCF.

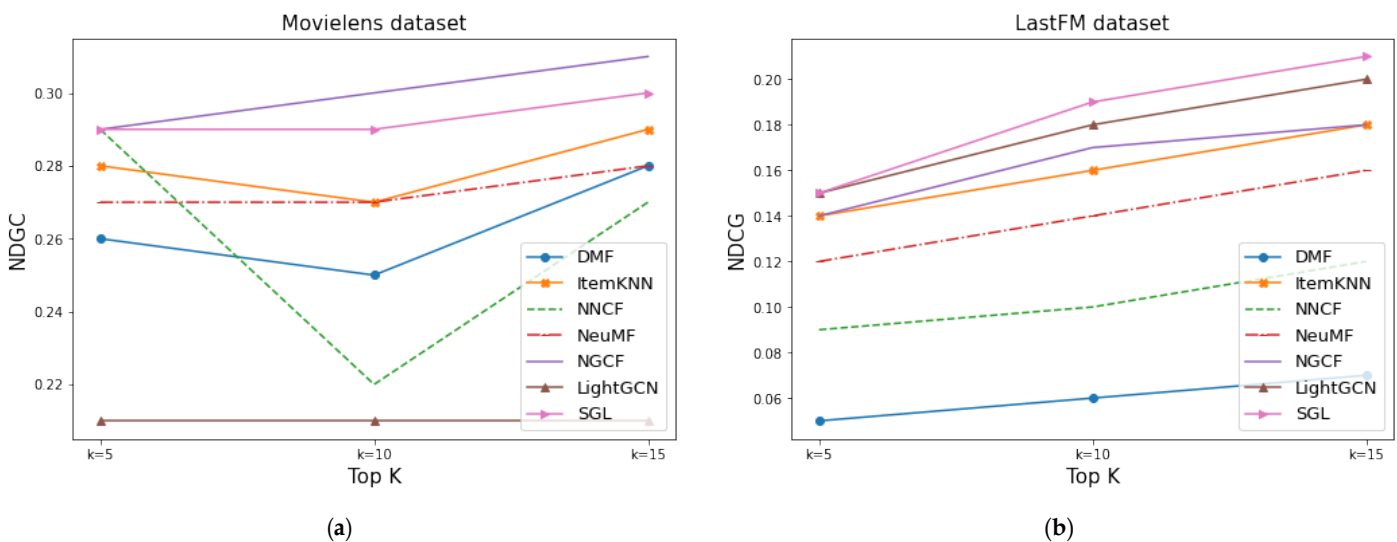

(**a**)            (**b**)

**Figure 15.** Results of NDGC for Movielens and LastFM. (**a**) MovieLens NDCG. (**b**) LastFM NDCG.

After presenting the results corresponding to the performance metrics, we turn to the bias metrics. Figure 16 illustrates the Gini Index of the given models. These results show that more accurate models such as SGL provided a lower Gini Index, which indicates lower diversity and more bias amplification. However, the differences in the results of the tested models were not as significant for this metric as for the previous metrics, especially in the case of the LastFM dataset.

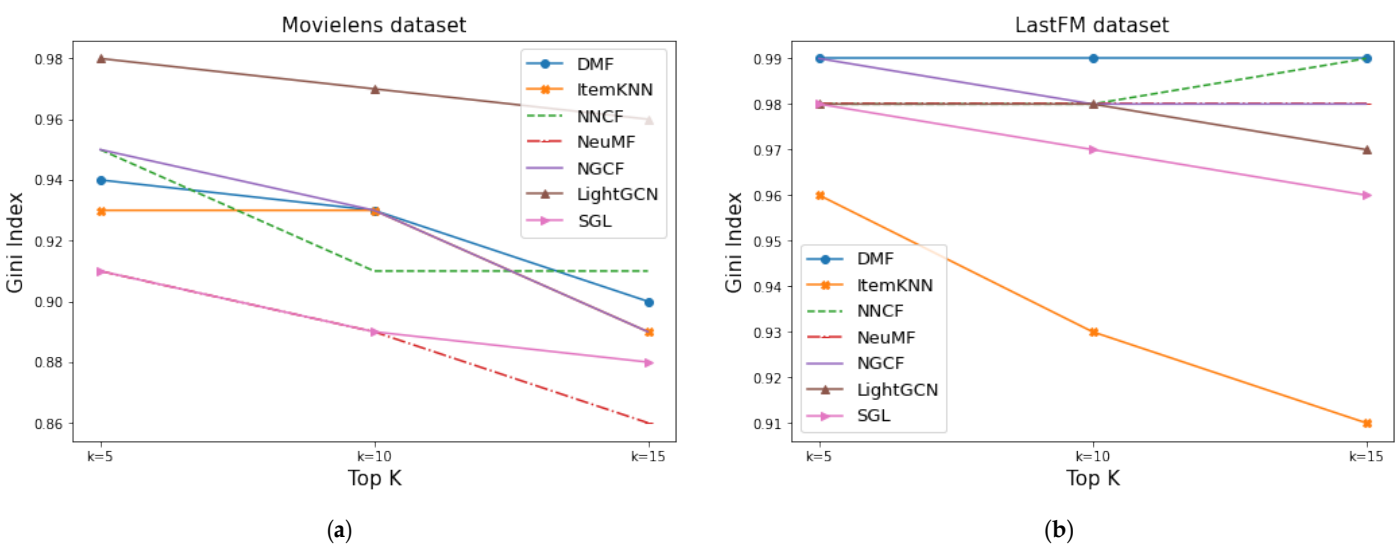

(**a**)            (**b**)

**Figure 16.** Results of Gini Index for Movielens and LastFM. (**a**) MovieLens Gini Index. (**b**) LastFM Gini Index.

Figure 17 shows item coverage. Low coverage represents discrimination with respect to certain items the user may like, but which are not recommended by the system. In general terms, the coverage was better for MovieLens than for LastFM, which is consistent with the greater number of items in the latter dataset, 41,269 items compared to 1682 in lastFM. It can be seen that NeuMF had the highest and LightGCN the lowest item coverage among all for the MovieLens dataset. In contrast, ItemKNN provided the best item coverage on the LastFM dataset. We can also see that the SGL method, based on the GNN, occupied the second position in the coverage ranking on both datasets, and the methods corresponding to the other approaches presented a different behavior for each dataset. NGCF coverage was similar to that of other non-GNN based methods on both datasets, and in the case of MovieLens, for k = 10 and K = 15, it outperformed most of those methods. This shows that the behavior of the methods based on the GNN in relation to coverage is acceptable.

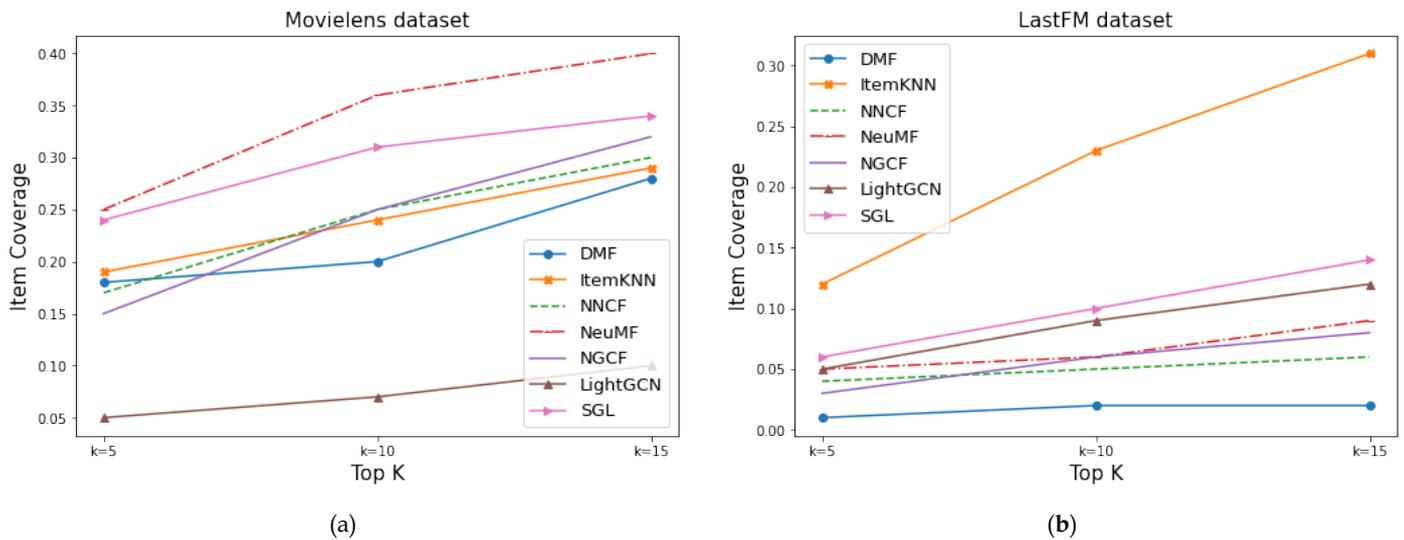

**Figure 17.** Results of item coverage for Movielens and LastFM. (**a**) MovieLens item coverage. (**b**) LastFM item coverage.

Figure 18 shows the average popularity of items on both datasets. Minimizing this bias involves generating recommendations of items with low popularity, so lower values of this metric would be the most desirable. Regarding these results, we can highlight the unequal behavior of the methods for the MovieLens dataset, since they present very different values between them and also present great variation for the different sizes of the top-K lists. The results showed that SGL algorithm provided the lowest values of popularity on both datasets and NeuMF gave very similar values on the MovieLens dataset. The worst performance was presented by LightGCN on the MovieLens dataset and NNCF on LastFM. NGCF also performed worse than most classical approaches on both datasets. Therefore, this confirms that there is an amplification in the popularity bias in the GNN-based methods, with the exception of SGL, which is superior to the rest regarding popularity and many of the other metrics.

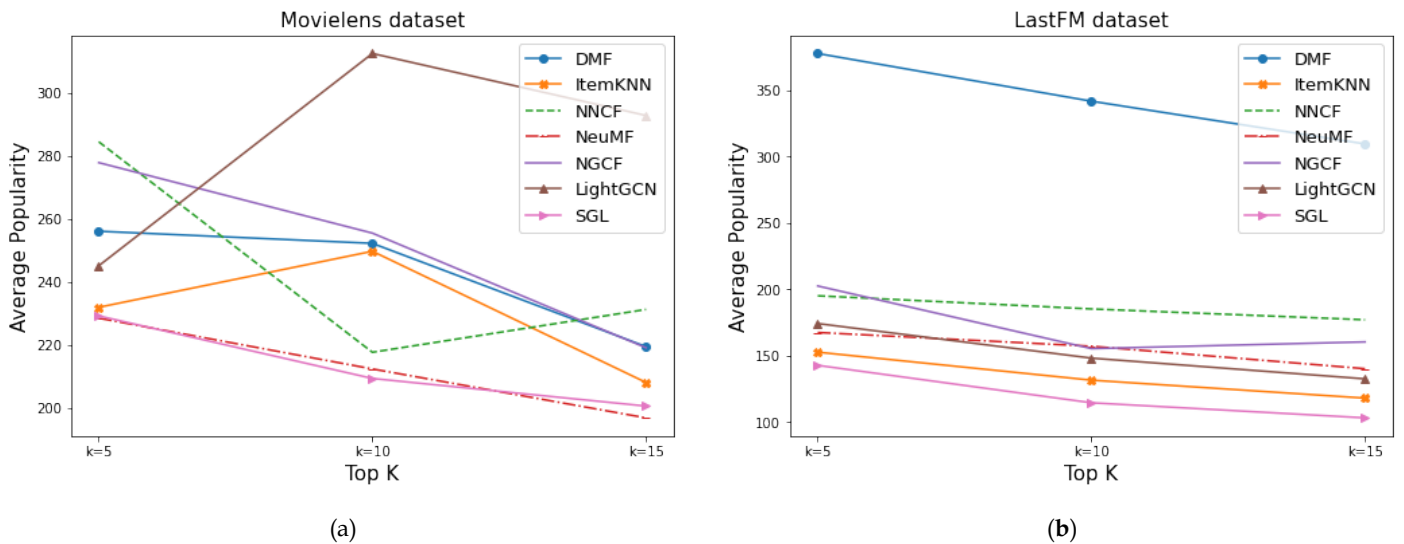

**Figure 18.** Results of average popularity for Movielens and LastFM. (**a**) MovieLens average popularity. (**b**) LastFM average popularity.

Figure 19 shows the results of differential fairness for the sensitive attribute gender. It should be taken into account that the lower its value, the lower the gender bias is. This metric applies to rating prediction since it uses the score prediction for user–item pairs and is not applicable to top-K recommendation lists. This is the reason why the figure does not show results for different values of k, but a single value corresponding to the mean values obtained for all the rating predictions for the examples in the test set. The graph shows that DMF on the LastFM dataset and ItemKNN on the Movielens dataset had a significant gender bias. Another relevant observation is the fact that the GNN-based methods provided better results than the rest in the case of the MovieLens dataset, while for LastFM, the values were more equal. In both cases, the SGL method outperformed the others, in line with the results discussed previously. Consequently, it follows that not only GNN-based methods are sensitive to gender bias, but also other methods of lower reliability. Even some models such as SGL performed better in comparison to the other methods used.

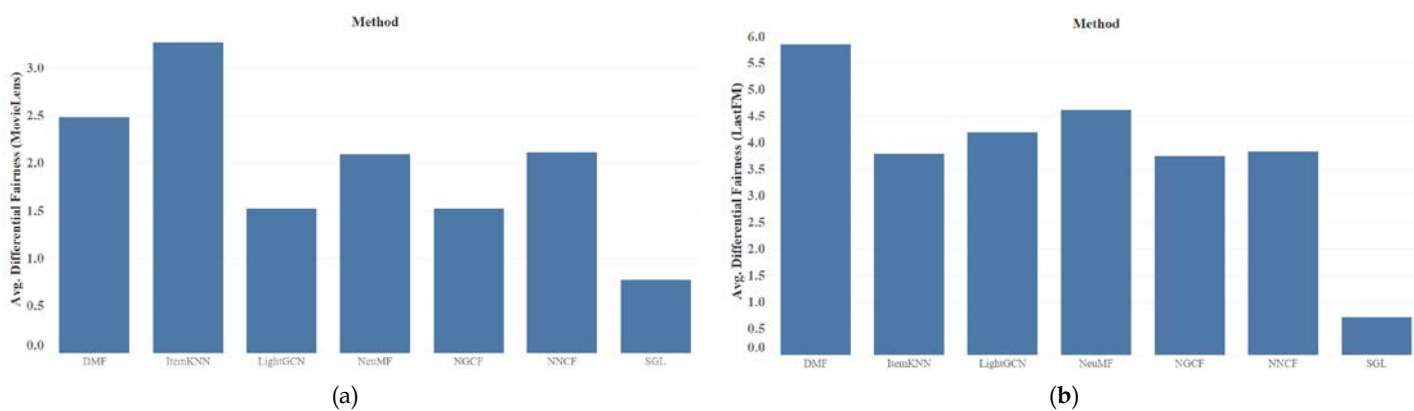

(a)

(b)

**Figure 19.** Results of differential fairness of sensitive attribute gender for Movielens and LastFM. (**a**) MovieLens average differential fairness of gender. (**b**) LastFM average differential fairness of gender.

Analyzing the results as a whole, it can be concluded that the GNN-based methods provide more reliable recommendations, since they achieved the highest values of the performance metrics. However, some types of biases were amplified when using these methods, as shown by the applied bias metrics, especially those related to diversity and coverage. Within this approach, the SGL algorithm showed the best performance since it had good reliability and was not as affected by biases as the other two algorithms in this group. The self-supervised learning on the user–item graph introduced in SGL to address some problems of recommenders based on graph convolution networks (sparsity, skewed distributions, and noise) also led to better performance against different types of biases, as evidenced by the results of this study. Additionally, this algorithm was the one that had the most uniform behavior for both datasets and for all list sizes. At the opposite extreme was LightGCN, which had a high sensitivity to some biases, and its results varied greatly from dataset to dataset.

## 5. Conclusions and Future Work

Bias is one of the most important issues in RSs, and due to its heavy costs for companies and individuals, this subject matter is worth investigating. In this work, we conducted an empirical study on bias amplification in GNN-based RSs. We studied the biases that are currently of most concern to companies that offer products or services to their users through the recommendation of top-K item lists. Moreover, we also implemented an investigation into gender bias by calculating differential fairness for the sensitive attribute gender. Apart from the fact that the lists should match the users' tastes as closely as possible, the most sought-after requirements of the lists are diversity, covering as many items as possible and containing unpopular items, in order to broaden and diversify the offer to users and help them discover products they would not otherwise be able to find.

To achieve our objectives, we selected two real-world datasets affected by biases, which were the subject of this research. The chosen datasets showed a power law distribution, also known as the long-tail phenomenon, which indicates popularity bias on items. Besides, the mentioned datasets contain the sensitive feature gender, which made them a good fit for this investigation.

Several different models from three approaches (CF, MF, and GNN) were implemented to compare the behavior of GNN-based methods against algorithms of other types. In order to evaluate the results, different metrics were used for measuring both the reliability and bias amplification of models for top $K$ recommendations. The metrics used to evaluate for each model the types of biases analyzed in this study were the Gini Index, item coverage, and average popularity.

The results showed that GNN-based methods mostly provide better performance regarding precision, recall, NDCG, and other reliability metrics, but are more prone to bias amplification based on the calculated bias metrics. Among all of them, LightGCN had the highest variability depending on the dataset and types of biases, while SGL was the most stable and the least sensitive to biases, especially popularity and coverage biases. This highlights the need for further research in order to achieve a better trade-off between accuracy and bias amplification. Furthermore, the results of gender bias showed a different pattern in bias amplification. In most cases, the GNN methods performed better with respect to differential fairness on both datasets. This shows that GNN approaches can have different amplification effects on different types of biases.

As future work, our purpose will be to investigate the causes of bias amplification in GNN-based recommendation algorithms and to propose mitigation solutions that minimally impact the reliability of recommendations. In addition, we intend to extend the study by making use of more datasets and a larger number of metrics, which will allow us to analyze sensitive features other than gender. Besides, an investigation can be implemented on gender bias with a non-binary gender attribute.

**Author Contributions:** Conceptualization, N.C., N.S. and M.N.M.-G.; methodology, N.C. and M.N.M.-G.; software, N.C. and N.S.; validation, N.C. and N.S.; formal analysis, N.C.; investigation, N.C. and N.S.; resources, M.N.M.-G.; data curation, N.C.; writing—original draft preparation, N.C.; writing—review and editing, N.C., N.S. and M.N.M-G; visualization, N.C.; supervision, M.N.M.-G.; project administration, M.N.M.-G.; funding acquisition, M.N.M.-G. All authors have read and agreed to the published version of the manuscript.

**Funding:** This research received no external funding.

**Informed Consent Statement:** Not applicable.

**Data Availability Statement:** Publicly available datasets were used. Details are provided in Section 3.1.

**Acknowledgments:** In this section you can acknowledge any support given which is not covered by the author contribution or funding sections. This may include administrative and technical support, or donations in kind (e.g., materials used for experiments).

**Conflicts of Interest:** The authors declare no conflict of interest.